# LEO-to-Ground Optical Communications using SOTA (Small Optical TrAnsponder)
## –Payload Verification Results and Experiments on Space Quantum Communications–

**Alberto Carrasco-Casado, Hideki Takenaka, Dimitar Kolev, Yasushi Munemasa, Hiroo Kunimori, Kenji Suzuki, Tetsuharu Fuse, Toshihiro Kubo-oka, Maki Akioka, Yoshisada Koyama, Morio Toyoshima**
National Institute of Information and Communications Technology (NICT)
Space Communications Laboratory, Wireless Network Research Center
4-2-1 Nukui-Kitamachi, Koganei, Tokyo, Japan, 184-8795
alberto@nict.go.jp

Free-space optical communications have held the promise of revolutionizing space communications for a long time. The benefits of increasing the bitrate while reducing the volume, mass and energy of the space terminals have attracted the attention of many researchers for a long time. In the last few years, more and more technology demonstrations have been taking place with participants from both the public and the private sector. The National Institute of Information and Communications Technology (NICT) in Japan has a long experience in this field. SOTA (Small Optical TrAnsponder) was the last NICT space lasercom mission, designed to demonstrate the potential of this technology applied to microsatellites. Since the beginning of SOTA mission in 2014, NICT regularly established communication using the Optical Ground Stations (OGS) located in the Headquarters at Koganei (Tokyo) to receive the SOTA signals, with over one hundred successful links. All the goals of the SOTA mission were fulfilled, including up to 10-Mbit/s downlinks using two different wavelengths and apertures, coarse and fine tracking of the OGS beacon, space-to-ground transmission of the on-board-camera images, experiments with different error correcting codes, interoperability with other international OGS, and experiments on quantum communications. The SOTA mission ended on November 2016, more than doubling the designed lifetime of 1-year. In this paper, the SOTA characteristics and basic operation are explained, along with the most relevant technological demonstrations.

## I. INTRODUCTION

In the last few years, free-space optical communications have gained significant importance as higher and higher amounts of data are being produced onboard all kinds of spacecraft. The potential of increasing the bitrate while reducing the volume, mass and energy of the space terminals, and taking advantage of a license-free spectrum [1] are attracting the attention of many participants from both the public and the private sector.

The long-awaited promise of revolutionizing space communications has started to materialize with an increasing number of projects all over the world. Lasercom demonstrations have been carried out in a wide variety of scenarios, e.g. deep space-to-ground [2], ground-to-deep space [3], LEO-to-ground [4][5], LEO-to-LEO [6], GEO-to-ground [7][8], GEO-to-LEO [9][10], GEO-to-aircraft [11], aircraft-to-ground [12][13], aircraft-to-aircraft [14] and balloon-to-balloon [15], and a number of projects are projected for future demonstrations [16][17][18][19][20][21].

The National Institute of Information and Communications Technology (NICT) in Japan has been engaged in research-and-development activities related to free-space optical communications for over 30 years. During this time, NICT has performed numerous demonstrations, including in-orbit validations such as ETS-VI in the 1990s [7] and OICETS in the 2000s [4]. The last one (in the 2010s) is the SOTA (Small Optical TrAnsponder) mission, conceived to prove for the first time the feasibility of high-bitrate lasercom from a microsatellite platform.

## II. SOTA MISSION

The maximum data rate achievable from space scales with the mass of the spacecraft. Even nowadays, the Gbps-class lasercom systems need very big space platforms, and the same thing happens with high-throughput RF systems. Small satellites have a strong limitation in the available bandwidth when using RF, as this technology is reaching its limits. For example, the maximum achievable bandwidth for cubesats is currently in the order of several tens of Mbps from LEO [22]. For the same kind of platform, data rates in the order of the Gbps will be achieved soon by lasercom [23] [18].

Furthermore, the crowded radio-frequency spectrum and regulatory difficulties to get the necessary authorizations are other issues that prevent many small satellites to get enough bandwidth, even if it is technically feasible using RF. Optical Communications do not require any authorization in terms of spectrum allocation, which is a big advantage, especially for small satellites with shorter development time frames.

In the late 2000s, NICT identified the potential of lasercom applied to this kind of satellites. With this idea in mind, SOTA was designed to demonstrate an optical communication system in a microsatellite platform for the first time. Furthermore, the goal was using COTS parts mainly, in order to demonstrate a simple yet effective communication system. This tendency has gained a great deal of interest in the last few years, taking the concept even further with the application of lasercom in cubesats platforms [24].

SOTA was embarked on SOCRATES (Space Optical Communications Research Advanced Technology Satellite), which was launched on May 24th, 2014, as a hosted payload on a Japanese H-II A rocket from the Tanegashima Space Center, in Japan. On July 11th, 2014, the first light of SOTA was received in the Optical Ground Station (OGS) of NICT Headquarters, in Koganei, Tokyo (Japan).

### III. SOTA SPECIFICATIONS

The mass of SOCRATES is 48 kg and its size 496×495×485 mm when the solar panels are folded. The mass of SOTA is only 5.9 kg and the maximum power consumption during operation is below 40 W. It is articulated using an alt-azimuth gimbal with micro-stepping motors with a reduction ratio of 100 to get a resolution of 20 µrad/pulse and a speed of 3°/s.

Fig. 1 shows the main elements of the receiving and the transmitting subsystems of SOTA. The receiving system is responsible of supporting the uplink and has two functions: ATP (Acquisition, Tracking and Pointing) and communications. ATP is performed in two levels: coarse (using a 2.3-cm lens with a FOV of 80 mrad) and fine (using a 5-cm Cassegrain telescope with a FOV of 4 mrad). The fine-detector provides a control signal for a fine-steering mirror in order to fine-point the transmitted communications laser.

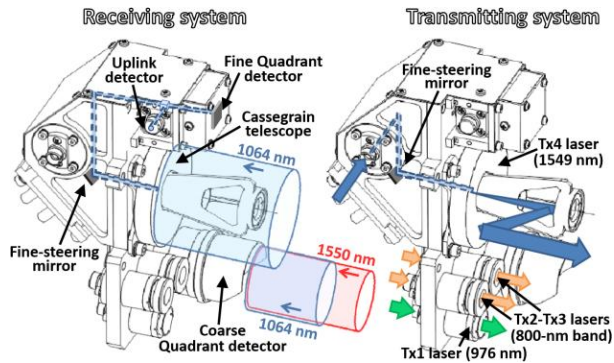

Fig. 1: Receiving and transmitting subsystems onboard SOTA including their main elements.

The transmitter subsystem is made up by four different lasers: Tx1 at 976 nm, Tx2 & Tx3 at 800-nm band and Tx4 at 1549 nm. Tx1 and Tx4 are communication lasers which can transmit at 1 or 10 Mbit/s with NRZ-OOK (Non-Return to Zero On-Off Keying) modulation from different selectable sources (an image from an onboard camera, a preloaded sample image or a pseudorandom binary sequence PRBS-15). This data can be coded against transmission errors using LDGM or Reed Solomon.

### IV. TYPICAL SOCRATES PASS

SOCRATES is inserted in a Sun-synchronous near-circular orbit at an altitude of ~600 km, with an inclination of 97.9° and a period of 97.4 minutes. In a typical SOCRATES pass, the sequence to establish a link goes in the following way: the experiment parameters are transmitted to the satellite by a RF link from a TT&C (Telemetry, Tracking and Control) ground station; the sequence of data to transmit through the optical link is prepared adding the error-correcting code; the OGS starts tracking the satellite according to the predicted orbital information and transmits a high-power beacon towards that position; when SOTA detects the beacon signal, it starts transmitting the laser or lasers set for that experiment until the end of the communications.

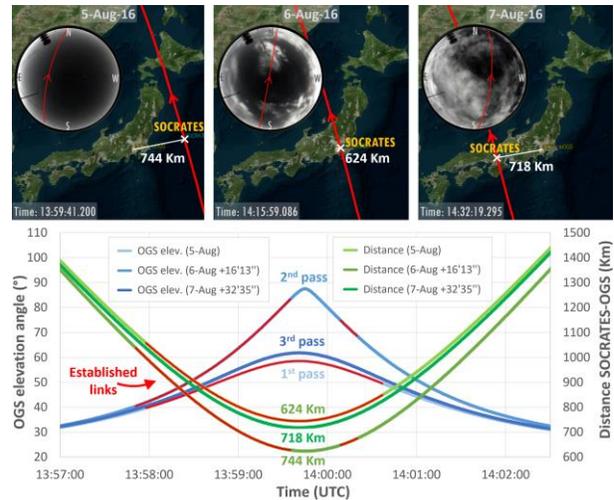

Fig. 2: An example of a typical 3-day sequence of SOCRATES passes. Above, a summary image of the three passes. Below, the variation of the OGS elevation angle and the distance SOCRATES-OGS for the three passes.

The characteristics of the orbit of SOCRATES allow the Koganei OGS to perform a lasercom experiment with SOTA in sequences of three days, with sequences of other three days without experiments. These links are always performed during the night around 23:00 (local time) because of SOTA design constraints. Fig. 2 shows an example of a typical three-day sequence of passes, including the satellite image of the passes (above) with the minimum SOCRATES-OGS distance (which happens at the maximum OGS elevation) and an image of the all-sky camera at that moment. Below, the figure shows the variation of the OGS elevation angle and the distance SOCRATES-OGS during these passes.

Generally, lasercom experiments occur when the OGS elevation angle is above ~30° and below ~80°. This can be observed in Fig. 2 for the links of August 5th and 6th in red colour. On August 7th, no experiment could be performed due to bad weather conditions. In these 3-day sequences, the intermediate pass requires an OGS high-elevation angle, which makes the link stop briefly until it is re-acquired by the ATP system.

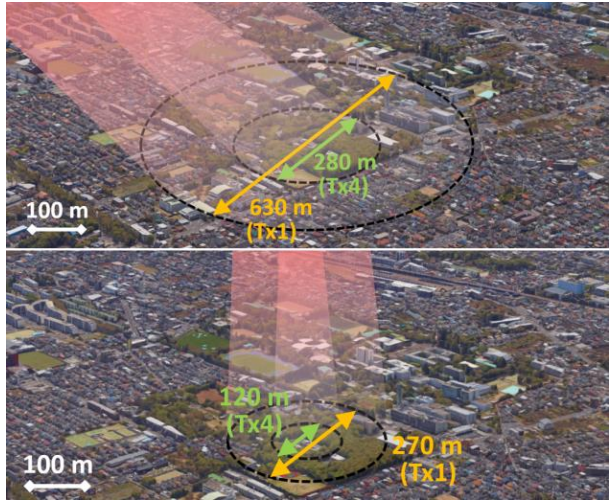

Fig. 3: Footprint of the SOTA lasers on the OGS of Koganei (Tokyo, Japan) for the minimum (~600 km, below) and maximum (~1400 km, above) distance range.

The different lasers of SOTA are designed to be transmitted with different divergences (see table 1). Tx4 is the main communication laser, with the bigger transmitting aperture and a fine-pointing subsystem, which makes it possible to transmit with the minimum divergence (223 µrad). Tx1 is the secondary communication laser and it is transmitted through a smaller aperture and without fine-pointing (although with a higher power), hence the divergence is bigger (double the Tx4 divergence). Tx2 and Tx3 are not communication lasers and their divergence is even bigger. Fig. 3 shows these differences in terms of the footprint on the OGS, for the minimum (~600 km) and maximum (~1400 km) distance range.

Table 1: Main specifications of SOTA Tx1 and Tx4.

|  | Tx1 | Tx4 |
| --- | --- | --- |
| Wavelength | 976 nm | 1549 nm |
| Transmitted power | 0.89 MW/sr | 0.57 MW/sr |
| Divergence angle | 500 µrad | 223 µrad |
| Aperture size | ~1 cm | ~5 cm |
| Polarization | Linear | Circular |
| Pointing loss | -3.4 dB | -5.7 dB |

The Fig. 4 shows an example of a SOTA downlink on December 9th 2015. This experiment was chosen for being a typical low-elevation pass (no interruption around the 90° elevation) with a clear sky (no clouds) and long duration (around 200 s). Tx4 (λ = 1549 nm) was used in this downlink at a 10-Mbit/s constant bitrate. The figure also shows the measured BER (Bit Error Rate) associated to the PRBS-15 transmitted sequence, with an average value of $6 \cdot 10^{-3}$, after FEC (Forward Error Correction) decoding using a Low-Density Generator-Matrix/Low-Density Parity Check (LDGM-LDPC) code. Since the SOTA receiver sensitivity is -55 dBm, a BER increase can be observed when the received power falls below the sensitivity. The decrease in the measured power around the middle of the experiment is associated with alignment issues related to the higher speed the telescope must move to track SOCRATES at higher elevation angles.

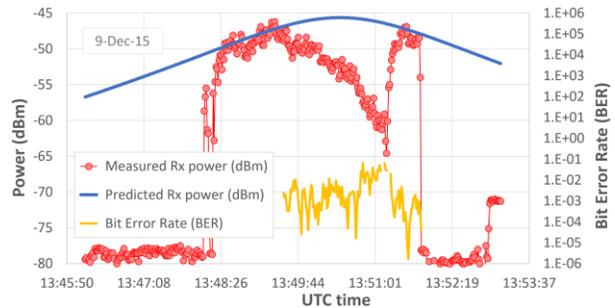

Fig. 4: Predicted and measured received power and BER measurement using the 1-m telescope in Koganei (Tokyo) on 9 Dec. 2015 in a Tx4 (1549 nm) downlink experiment at 10 Mbit/s.

In the link budget calculation, the atmospheric loss was simulated with MODTRAN, and it depends on the telescope elevation angle (the optical path is longer at lower elevation angles), varying between 3.2 dB at the longest distance (1300 km, at a 25° elevation angle) and 1.8 dB at the shortest distance (875 km, at a 42° elevation angle). The elevation angle also determines the free-space loss, as the optical path is longer for lower elevation angles, producing a bigger footprint, as was shown in the Fig. 3. The beam scintillation, i.e. the power fluctuation due to atmospheric turbulence, depends on the telescope elevation angle as well. An in-depth analysis of the power scintillation using SOTA measurements can be found at [25].

## V. SOTA ACHIEVEMENTS

SOCRATES was launched on May 24th, 2014. A general verification of the satellite was performed during ~2 months, followed by a start-up period of the OGS and SOTA. The lasercom experiments officially started in November, 2014, and they have continuously been performed up to the present date. With nearly 300 planned passes, around 45% of the experiments were carried out. The rest have been cancelled due to bad-weather forecast or due to maintenance work on the satellite or the lasercom terminal. Almost 30% of the performed experiments were carried out successfully with a stablished uplink and downlink, and the rest were unsuccessful due to bad weather or abnormal operation. The SOTA mission ended on November 2016 [26], doubling the 1-year initial lifetime design.

In SOTA mission, several success criteria are defined. The minimum success was achieved after the verification of the correct behavior of all the components of SOTA. The success level consisted of the verification of the ATP system after receiving the uplink beacon and BER measurements of the Tx1 and Tx4 signals in the OGS. Full success was achieved when the different data sources (the camera image, the preloaded image and the pseudorandom binary sequence) could be transmitted to the OGS, including error correcting codification (see Fig. 5).

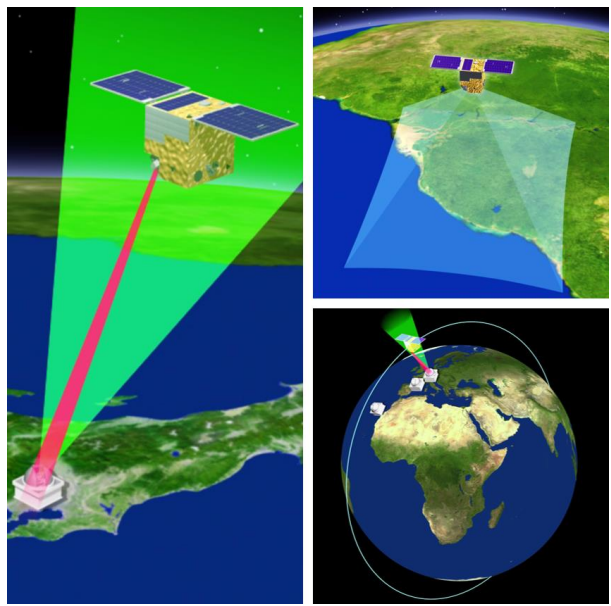

Fig. 5: SOCRATES artistic illustrations. Left: Lasercom between SOTA and NICT OGS in Koganei, Tokyo. Top right: SOCRATES imaging the Earth with a visible camera. Bottom right: International campaign with other ground stations (in the image, MeO OGS, France).

An extra-success phase was carried out since May 2015, consisting of interoperability tests with other international OGSs (see Fig. 6). Four different space agencies participated in the campaign: ESA (European Space Agency), CNES (National Centre for Space Studies), DLR (German Aerospace Center) and CSA (Canadian Space Agency). Two passes were assigned to ESA in April 2015 using the 1-m OGS in Tenerife (Spain), but the uplink beacon hit SOTA only for a short time and with low power, thus no link was established. CNES performed five successful links detecting Tx1 and Tx4 signals from SOTA during June, July and October 2015 using the 1.54-m MeO OGS in Caussol (France) [267]. DLR hit SOTA with the uplink beacon in July 2015 and they performed one successful link in May 2016 receiving the Tx4 signal from SOTA using a 40-cm and 60-cm OGS in Oberpfaffenhofen (Germany) [28]. Lastly, CSA used a 40-cm OGS in Montreal (Canada) to perform a partially-successful link in August 2016, hitting SOTA with the uplink beacon in one pass.

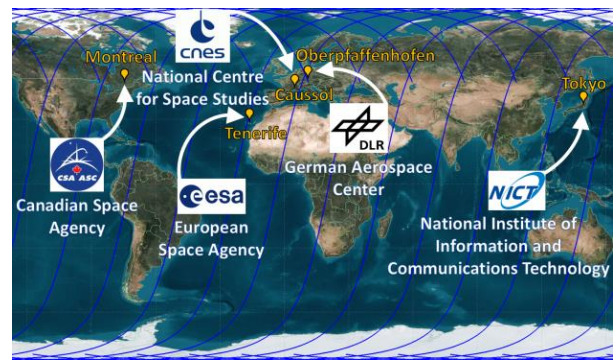

Fig. 6: SOTA interoperability campaign: the international OGSs which participated in the campaign are shown.

## VI. SITE DIVERSITY

The main impediment of free-space optical communications is the propagation of laser signals through the atmosphere. An effective solution to mitigate this problem is site diversity. This technique consists of replicating the ground segments in several uncorrelated sites in order to increase the probability of link success. Within this endeavor, NICT is developing a network of OGSs throughout Japan (Fig. 7), with autonomous-operation capabilities in order to be operated remotely from a control center according to the link availability, decided using meteorological data in each location.

This network is called INNOVA (IN-orbit and Networked Optical ground stations experimental Verification Advanced testbed) [29] and it consists of three different OGS locations: NICT Headquarters (Koganei, Tokyo prefecture), Kashima Space Technology Center (Kashima, Ibaraki prefecture) and Okinawa Electromagnetic Technology Center (Okinawa, Okinawa prefecture). All these OGSs include a telescope with a 1-m aperture and a focal length of 12 m, with LEO-tracking capabilities (accuracy better than 10 arcsec), and five available focus (Cassegrain, 2×Nasmyth and 2×Coudé) in the case of Koganei and one Nasmyth focus in the case of Kashima and Okinawa.

INNOVA also includes a total number of ten locations with different meteorological sensors to take long-term weather statistics.

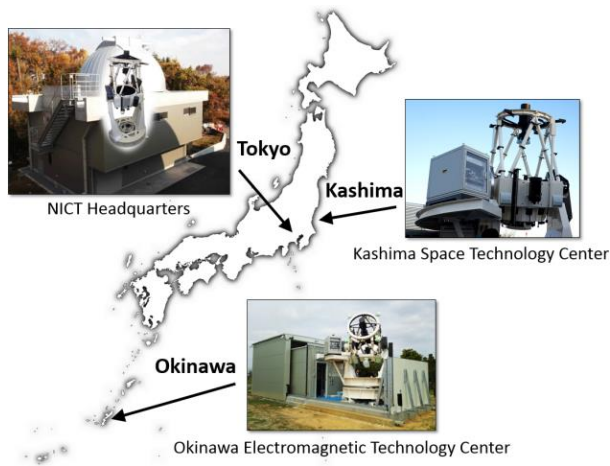

Fig. 7: Map of Japan with the three NICT 1 m-class Optical Ground Stations for site diversity.

As an example of the potential improvement of the site diversity, an estimation of the link availability was made taking into account all the possible SOCRATES passes over the three NICT OGS locations during the period of June 5th to July 18th. During the SOTA mission, the Kashima and Okinawa OGSs were still not in operation, thus only the Koganei OGS was used for SOTA experiments. The following study presents a theoretical estimation using the SOTA mission as a study case assuming that all the NICT OGSs could have been used during the mission.

Table 2: Statistics of the possible passes over the three NICT OGSs during the 2016 rainy season.

|  | Koganei | Kashima | Okinawa |
| --- | --- | --- | --- |
| Period (Baiu) | June 5th to July 18th | | |
| No. of days | 44 | | |
| Possible passes | 21 | 21 | 21 |
| Total link time | 7156 s | 7161 s | 7143 s |
| Link probability | 50% | 49% | 31% |
| Real link time | 3607 s | 3473 s | 2193 s |
| Using 2 OGS | 5508 s (77%) | | - |
| Using 3 OGS | 7174 s (100%) | | |

According to the Japan Meteorological Agency, the 2016 rainy season occurred in this period in Kanto prefecture, where the NICT headquarters is located with the Koganei OGS. All of Japan (except Hokkaido island) experiences this rainy season, called Tsuyu, every year in early summer (from early June to late July, and one month earlier in Okinawa). Although Tokyo registers an average of 12 rainy days during the whole rainy season, just 120 hours of sunshine are recorded on average. The following figures illustrate the strong impact this cloudy season has on the link availability: if the annual clear-sky rate in Koganei OGS is 56.5%, the one for the 2016 rainy season was only 11.2%. Therefore, this period can be used as a worst-case scenario in terms of link availability.

INNOVA meteorological data can be used to predict the OGS availability for lasercom. A simple predictive algorithm was designed taking into account the atmospheric transmittance and the cloud coverage. The atmospheric transmittance is estimated using the solar radiation, and the cloud coverage is estimated using temperature measurements and all-sky camera images. In general, the algorithm predicts good conditions for lasercom if atmospheric transmittance is over 60% and the cloud coverage is less than 10%. The current predictive algorithm is in a preliminary version of this on-going project.

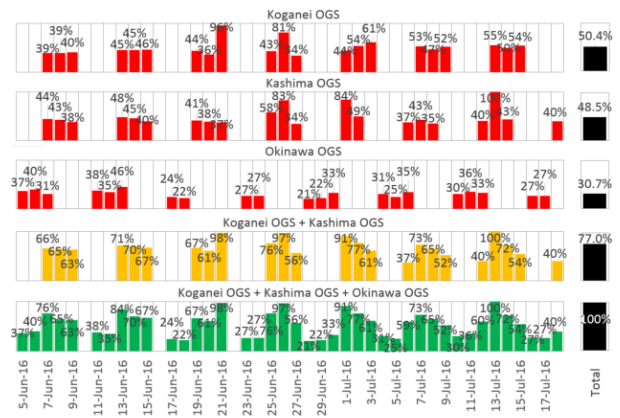

Fig. 8: Link probability for each SOTA pass in the 2016 rainy season using each single NICT OGS (red) and applying site diversity with 2 OGSs (orange) and 3 OGSs (green).

During the 2016 rainy season, each OGS had 21 available passes according to the usual constraints for SOTA experiments: the experiment has to be at nighttime and at an OGS elevation over 45°. The table 2 shows the statistics for this period. If only one OGS was used, the best link probability would be 50% (at Koganei). Assuming a weather decorrelation between the three sites, if two OGS (Koganei or Kashima) could be selected in each pass, the link probability would increase up to 77%, and if the three OGS could be selected, the link probability would become 82% on average in every SOTA pass.

The previous calculations are average values assuming that each pass occurs on the same day and one out of two or three OGS can be selected in each pass to avoid the cloud coverage. This is a realistic assumption in the case of selecting between Koganei and Kashima because the SOTA passes occur almost at the same time. As shown in the Fig. 8, for Okinawa the SOTA passes usually occur on different days, thus different passes than the Koganei/Kashima ones. Therefore, the combined link probability for 2 OGSs was calculated only for Koganei and Kashima.

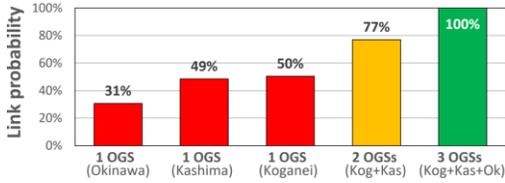

Fig. 9: Individual link probability for every NICT OGS, and link probability applying site diversity with 2 OGSs (orange) and 3 OGSs (green).

The fact that the Okinawa passes ocurr on different days than the Koganei/Kashima ones in practice increases the total available link time, reaching the 100% of the foreseen link time for one OGS. This means that even in the rainy season (worst-case scenario for Japan) the three NICT OGSs allow to download the same data as having a single OGS with a cloud coverage of 0%. The link probabilities for each individual OGS as well as the probabilities for 2-OGSs and 3-OGSs site diversity are summarized in the Fig. 9.

## VII. QKD EXPERIMENT

Free-space optical communications can allow other applications besides the high-speed transmission of data from a satellite. One important application is the information security, and specifically quantum communication can be carried out using lasercom terminals with some modifications. One of the most important applications of quantum communication is Quantum Key Distribution (QKD). This technique allows the transmission of encryption keys in a theoretically secure way by using fundamental laws of physics, in particular the Heisenberg uncertainty principle. QKD is a relatively-mature technique in its guided version, using optical fibers. However, the transmission distance is limited to ~200 km [30] because of the lack of reliable quantum repeaters. Space-to-ground links, being a key step towards a global QKD network remain to be demonstrated [31].

A basic QKD experiment was designed as part of the extra-success phase of the SOTA mission. This experiment makes use of the Tx2 and Tx3 lasers (see Fig.1) in the 800-nm band. They are non-orthogonally linearly polarized laser sources, since the most important QKD protocols are based on polarization-encoded photons. In the first phase of this experiment, a characterization of the polarization behavior after the atmospheric propagation was carried out [32]. For this, a 1.5-m Cassegrain telescope was used to gather the signal from SOTA and couple it to a polarimeter (Fig. 10, left). These experiments were performed from January to March 2016 measuring the polarization of the Tx1 ($\lambda$ = 976 nm, linear polarization) and Tx4 ($\lambda$ = 1549 nm, circular polarization) signals when transmitting a PRBS-15 sequence at 10 Mbit/s.

The Stokes parameters, the received optical power and the Degree Of Polarization (DOP) were measured in these experiments. The polarimeter system has a strong dependence between the DOP and the received power: If the power is high enough, a DOP = 100% can be measured from a well-polarized signal, but for low power, a decrease in the DOP as well as a bigger variability in the measurement is observed for the same signal. In the SOTA experiments, the power is not high enough for the polarimeter to measure the DOP with high accuracy, so the characterization of this dependence is important, and it was explained in detail in [32].

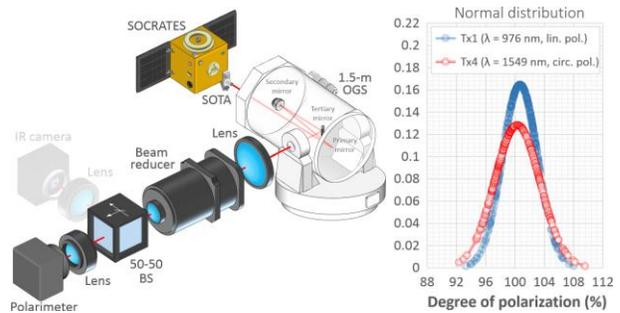

Fig. 10: Left: Receiving system for measuring the polarization of SOTA laser sources. Right: Normal distribution of the degree of polarization received in the NICT OGS from SOTA Tx1 ($\lambda$ = 976 nm, linear polarization) in blue, and Tx4 ($\lambda$ = 1549 nm, circular polarization) in red.

Fig. 10 shows on the right side the normal distribution of the DOP measurements from SOTA Tx1 and Tx4 signals. For this figure, only powers over -55 dBm were selected (since below this level, the DOP decreases rapidly). However, such a low power adds some uncertainty in the measurement, and the DOP can reach impossible values above 100%. The standard deviation associated to this fit agrees well with the expected one for this power range according to the characterization of the measurement system, hence the average value for DOP ≈ 100% observed in Tx1 and Tx4 is a reliable measurement. This allowed to conclude that the linear polarization, which is the basic property for encoding information in free-space QKD, is well preserved after the atmospheric propagation. In 2009, NICT measured the influence of the atmosphere in the LEO-ground propagation of the circularly-polarized 815-nm laser from OICETS [33]. The experiment with SOTA was the first time the linear polarization of a source in space was measured, as well as the polarization of a source at $\lambda = 1549$ nm. This wavelength and polarization are good candidates to become the standard in future QKD links from space: this spectral region is the preferred in terms of atmospheric transmission and turbulence, and linear polarization is used in the most extended protocols, such as BB84.

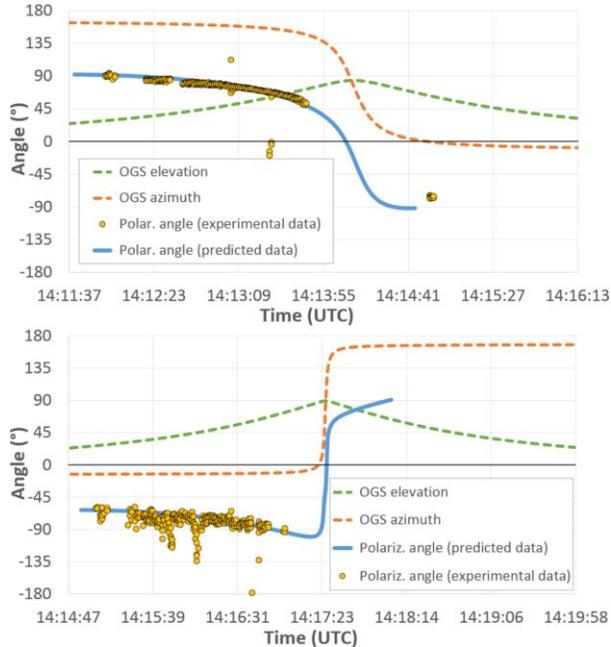

Fig. 11: Polarization angle received in the NICT OGS from SOTA Tx1 ($\lambda = 980$ nm, linear polarization) above, and Tx4 ($\lambda = 1550$ nm, circular polarization) below.

After the verification of the polarization preservation through the atmosphere, other experiment was performed, consisting of demonstrating for the first time some basic principles of quantum communication from space. Currently, the data obtained in this experiment is being analysed and the results will be reported soon. In this experiment, the non-orthogonal Tx2 and Tx3 signals were received at a near single-photon regime using a QKD-like receiver in the 1-m OGS. For this experiment to be accomplished successfully, a key step was being able to track and correct the polarization of the received signals due to the motion of the satellite in relation to the OGS. This motion makes the reference frame change with time, which has to be aligned. This can be achieved with a rotating half-wave plate before the receiver. Since the orbital information of the satellite is well known, it is possible to predict the angle of the linear polarization when received in the OGS. A simulation was carried out taking into account the relative motion between the SOCRATES satellite and the OGS, as well as the OGS elevation (which also makes the angle change when using the Nasmyth bench of the telescope). Fig. 11 shows this prediction for previous SOTA passes and a good agreement is observed with measured data using the polarimeter for the same passes. The two examples of the Fig. 11 show the prediction for Tx1 as well as Tx4 (the polarization angle for a circular polarization like Tx4 is possible to be measured since the signal is not perfectly circular, but elliptical). For both experiments, an interruption of the reception can be observed for high elevation angles, as was explained in the section IV.

## VIII. CONCLUSION

The SOTA lasercom terminal onboard the SOCRATES satellite was operated by NICT for more than two years. During this time, all the goals of the mission were achieved, including up to 10-Mbit/s downlinks using two different wavelengths and apertures, verification of coarse and fine tracking of the OGS beacon, space-to-ground transmission of pseudo-random sequences and images from the on-board-camera as well as preloaded samples, and experiments with different error correcting codes. Other two extra-success experiments were carried out within SOTA mission: interoperability with other international OGS and basic experiments on space QKD. In this paper, the basic characteristics of SOTA and the fundamentals of its operation were described, along with an overview of the achievements of the mission.